\documentclass[sigconf]{acmart}

\usepackage{graphicx}
\usepackage{afterpage}
\usepackage{graphicx}
\usepackage{textcomp}
\usepackage{xcolor}
\usepackage{enumitem}
\usepackage{hyperref}

\usepackage{multicol}
\usepackage[many]{tcolorbox}
\usepackage{wrapfig}
\usepackage{url}
\usepackage{float}
\usepackage{listings}
\usepackage[labelsep=colon,font=footnotesize]{caption}
\usepackage{xspace}
\usepackage{caption}
\usepackage{subcaption}
\usepackage{tikz}
\usetikzlibrary{svg.path}
\usepackage{multirow}
\usepackage{colortbl}

\usepackage{xcolor} 

\AtBeginDocument{%
  }

\copyrightyear{2026}
\acmYear{2026}
\setcopyright{cc}
\setcctype{by}
\acmConference[FORGE '26]{2026 IEEE/ACM Third International Conference on AI Foundation Models and Software Engineering}{April 12--13, 2026}{Rio de Janeiro, Brazil}
\acmBooktitle{2026 IEEE/ACM Third International Conference on AI Foundation Models and Software Engineering (FORGE '26), April 12--13, 2026, Rio de Janeiro, Brazil}
\acmPrice{}
\acmDOI{10.1145/3793655.3793709}
\acmISBN{979-8-4007-2477-0/2026/04}

\begin{document}

\title{Tricky²: Towards a Benchmark for Evaluating Human and LLM Error Interactions}

\author{Cole Granger}
\affiliation{%
  \institution{William and Mary}
  \city{Williamsburg}
  \state{VA}
  \country{USA}
}
\email{cjgranger@wm.edu}

\author{Dipin Khati}
\affiliation{%
  \institution{William and Mary}
  \city{Williamsburg}
  \state{VA}
  \country{USA}
}
\email{dkhati@wm.edu}

\author{Daniel Rodriguez-Cardenas}
\affiliation{%
  \institution{William and Mary}
  \city{Williamsburg}
  \state{VA}
  \country{USA}
}
\email{dhrodriguezcar@wm.edu}

\author{Denys Poshyvanyk}
\affiliation{%
  \institution{William and Mary}
  \city{Williamsburg}
  \state{VA}
  \country{USA}
}
\email{dposhyvanyk@wm.edu}



\begin{abstract}
Large language models (LLMs) are increasingly integrated into software development workflows, yet they often introduce subtle logic or data-misuse errors that differ from human bugs. To study how these two error types interact, we construct \textit{Tricky\textsuperscript{2}}, a hybrid dataset that augments the existing \textit{TrickyBugs} \cite{TrickyBugs} corpus of human-written defects with errors injected by both GPT-5 and OpenAI-oss-20b across C++, Python, and Java programs. Our approach uses a taxonomy-guided prompting framework to generate machine-originated bugs while preserving original human defects and program structure. The resulting corpus spans human-only, LLM-only, and human+LLM splits, enabling analysis of mixed-origin error behavior, multi-bug repair robustness, and reliability in hybrid human–machine code. This paper outlines the dataset construction pipeline and illustrates its use through small-scale baseline evaluations of classification, localization, and repair tasks.
\end{abstract}

\ccsdesc[500]{Software and its engineering~Software testing and debugging}
\ccsdesc[500]{Computing methodologies~Machine learning}
\ccsdesc[500]{Computing methodologies~Natural language generation}
\ccsdesc[500]{Software and its engineering~Software reliability}

\keywords{Large language models, software testing, program repair, debugging, mixed-origin errors, machine learning for code}


\maketitle

\begin{figure*}[!t]
  \centering
  \includegraphics[width=\textwidth]{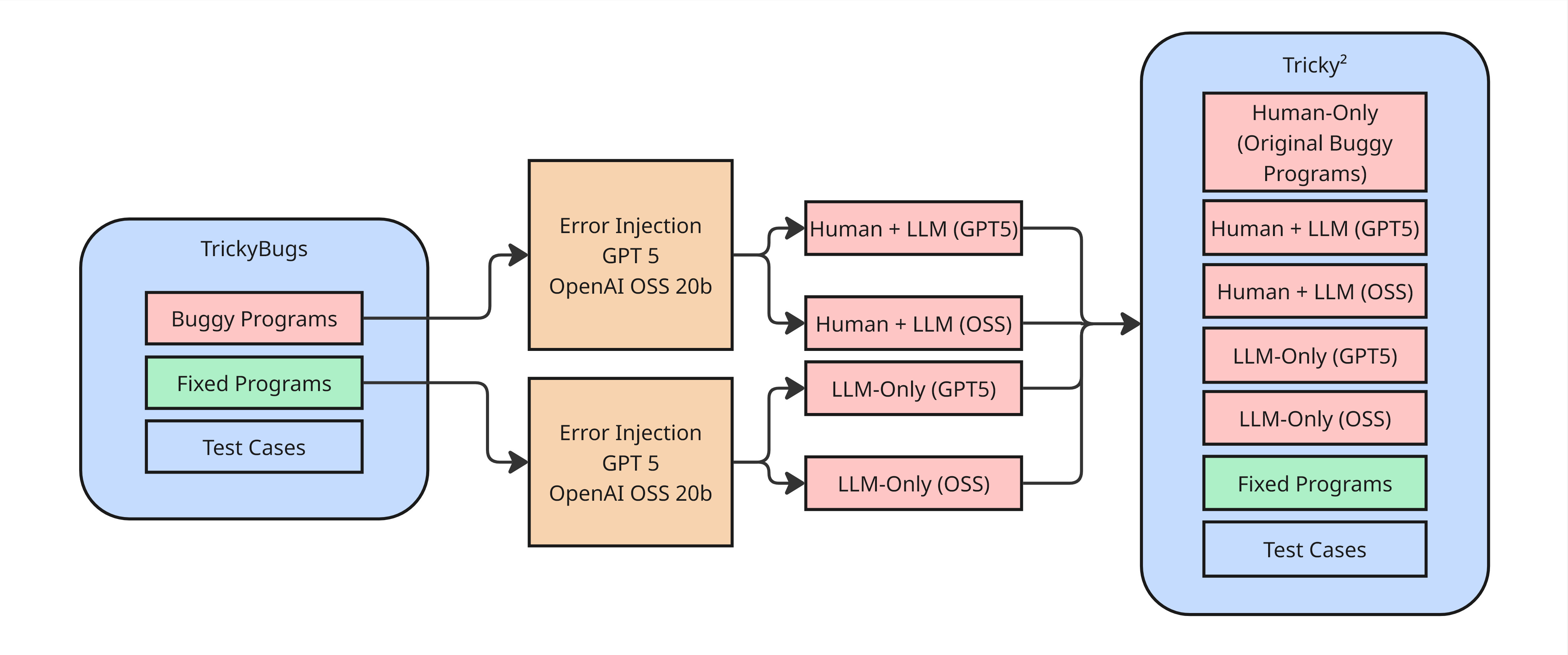}
  \Description{A flowchart with three branches showing how TrickyBugs data flows into three splits: Human-only from buggy submissions, Human+LLM from error-injected buggy programs, and LLM-only from error-injected fixed programs. All branches converge on shared test cases.}
  \caption{
  Construction of the \textit{Tricky\textsuperscript{2}} benchmark from the \textit{TrickyBugs} dataset.
  Buggy human submissions form the \textbf{Human-only} split,
  error-injected buggy programs yield the \textbf{Human+LLM} split,
  and error-injected fixed programs yield the \textbf{LLM-only} split.
  Each injection applies a controlled taxonomy.
  All splits share identical test cases and fixed references.
  }
  \label{fig:tricky2_pipeline}
\end{figure*}

\section{Introduction}

Large language models (LLMs) are transforming software engineering (SE) by enabling automated code completion, documentation \cite{AI_SE_Review, LLM_usage}, and error correction across languages, with tools like GitHub Copilot integrated into millions of developer workflows \cite{github_copilot_users, anthropic_code_use}. Although this collaboration accelerates development, it also introduces new reliability and security concerns \cite{copilot_security, basic2025vulnerabilitiesremediationsystematicliterature}. 

However, despite the rapid adoption of LLMs in programming tasks, existing benchmarks consider human- and AI-generated bugs in isolation. We\textit{BugsInPy}\cite{BugsInPy}, and \textit{TrickyBugs}\cite{TrickyBugs} capture real human defects, while synthetic collections like \textit{buggy-HumanEval} \cite{buggy-HumanEval} or CriticGPT \cite{CriticGPT} focus on LLM-generated errors. Although both have proven valuable, they evaluate human and AI errors in isolation. Recent works show that LLMs and human developers introduce qualitatively distinct defects \cite{wang2025understandingcharacteristicscodegeneration, tambon2024bugslargelanguagemodels}; AI code tends to be simpler, but more prone to unused structures, hallucinations, and high-risk security vulnerabilities \cite{copilot_security, basic2025vulnerabilitiesremediationsystematicliterature}, while human written code shows greater structural complexity and maintainability difficulty \cite{AI_SE_Review, characteristics_of_chatgpt}. Consequently, existing benchmarks cannot evaluate \emph{error-interaction}, or how human and AI bugs coexist, compound, or mask one another within the same code base. Current datasets capture human or AI behavior \cite{buggy-HumanEval, TrickyBugs, Defects4J, BugsInPy, ManyBugs, SWE-bench}, and not their interplay, leaving unanswered questions about debugging robustness, repair reliability, and how current evaluation metrics should adapt when multiple error sources are present. Evaluating such mixed-origin error settings has become increasingly important as real software development workflows now routinely combine human-written and LLM-generated code within the same files, functions, and pull requests. Developers often accept or partially modify LLM suggestions \cite{usabilityofcodegen}, creating hybrid contexts where human and AI bugs can interact in ways that make debugging more difficult—human fixes can mask LLM faults, LLM repairs can reintroduce human defects, and standard evaluation metrics can fail when multiple error sources are combined. Without benchmarks that model this co-occurrence, the SE community lacks the ability to study these interaction effects, rigorously evaluate repair tools in realistic mixed settings, or design methods for safe human–AI collaboration.

Based on this gap, we construct two research questions:
\begin{itemize}
    \item \textbf{RQ1:} Can mixed-origin error datasets reveal interaction effects that affect model repair performance?
    \item \textbf{RQ2:} How reliably can current LLMs classify, localize, and repair errors in hybrid human+AI code contexts?
\end{itemize}

To address this, we introduce \textbf{Tricky\textsuperscript{2}}, a benchmark that unifies human errors and LLM-originating errors within shared-program contexts. Unlike prior benchmarks that isolate error origins, Tricky\textsuperscript{2} explicitly models error co-occurrence, enabling controlled experiments on robustness, explainability, and repair across mixed human–AI code. Building on the \textit{TrickyBugs} dataset of real competitive-programming submissions \cite{TrickyBugs}, we employ GPT-5 and OpenAI-oss-20b to inject additional errors following a controlled taxonomy of error types---\textit{Input/Output}, \textit{Variable/Data}, \textit{Logic/Condition}, \textit{Loop/Iteration}, and \textit{Function/Procedure} \cite{taxonomy}. 
The resulting corpus spans C++, Java, and Python, organized into three splits:
\emph{human-only}, \emph{LLM-only}, and \emph{human+LLM}.
To make this setting empirically measurable, we define three complementary evaluation tasks capturing classification, error identification, and repair dimensions of reliability.
\begin{itemize}
    \item (1) \textbf{Origin Classification}: distinguishing human, LLM, and mixed defects; 
    \item (2) \textbf{Error Identification}: localizing bug spans and identifying taxonomy level; and 
    \item (3) \textbf{Program Repair}: evaluate the minimal patch success in the provided test cases.
\end{itemize}
This benchmark represents an early step toward evaluating the reliability of software in hybrid human–AI development environments. We envision Tricky\textsuperscript{2} as a foundation for future studies on agentic debugging and error-aware collaboration between developers and large language models.

\section{Methodology}
 
To study how human and machine-generated errors coexist, we extend \textit{TrickyBugs} \cite{TrickyBugs} with LLM-injected errors. \textit{TrickyBugs} contains 3,043 human-written buggy programs across 324 tasks and 1,361 fixed programs across 224 tasks in C++, Python, and Java. \textit{Tricky\textsuperscript{2}}, our extension, introduces errors from GPT-5 and OpenAI-oss-20b while preserving original human defects, enabling controlled investigation of error interaction and repair robustness in hybrid code settings.

\subsection{Prompt Design for Bug Injection}

\begin{figure*}[t]
  \centering
  \includegraphics[width=\textwidth]{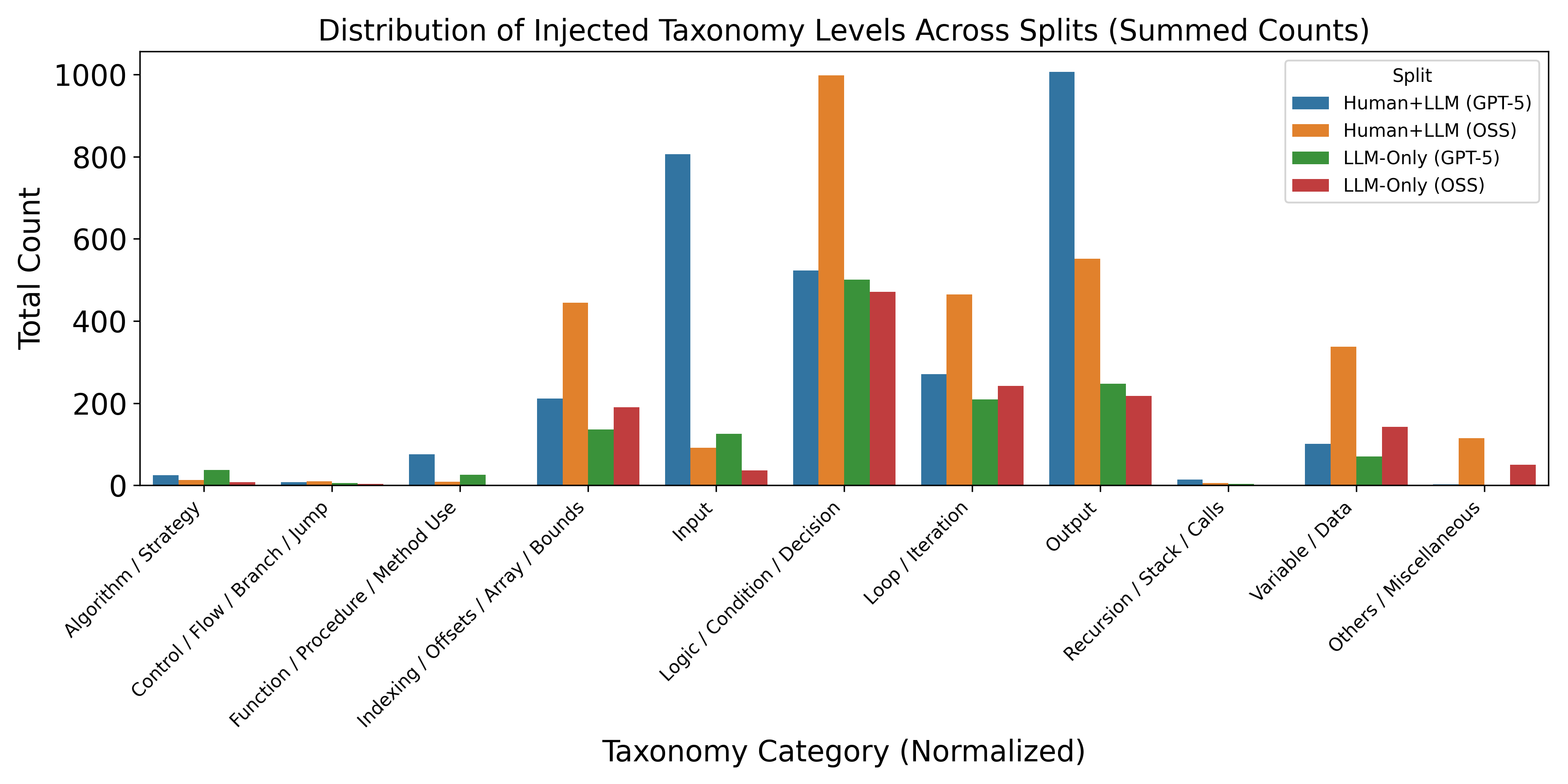}
  \Description{A grouped bar chart comparing the distribution of injected errors across five taxonomy categories (Input/Output, Variable/Data, Logic/Condition, Loop/Iteration, and Function/Procedure) for Human+LLM and LLM-Only splits. Each category shows two bars representing the two splits, with Human+LLM consistently having higher counts due to more programs in that split. The distribution pattern across taxonomy categories is similar between splits.}
  \caption{
  \textit{Distribution of the injected errors by taxonomy level and split (Human+LLM vs. LLM-Only).} There are fewer problems in the LLM-Only versions because there are not fixed programs for each buggy program in the original dataset. The taxonomy was provided with each injection prompt, but total numbers of each category were not provided to the model. 
  }
  \label{fig:tricky2_taxonomy_origins}
\end{figure*}

We designed a structured natural-language prompt that specifies the role of the LLM and behavioral constraints. The template defines the model's role ("an expert software engineer"), the allowable transformations, the bug taxonomy and level definitions, the required output format, and behavioral constraints. The same prompt template and constraints were applied to both GPT-5 and OpenAI-oss-20b to ensure comparable bug-injection behavior across models. The prompt instructs the model to: (1) inject exactly one new bug from a predefined taxonomy [2], (2) preserve all existing human bugs, (3) avoid comments or formatting changes, and (4) output the modified file with taxonomy level and character positions of the injected code.

The prompt contains a dedicated placeholder for the input program (delimited with code fences). We do not use multi-step model interactions or chained prompting. Each injection is produced by one call to a single model with the same template. To support reproducibility, we release the full prompt template. Although naturally occurring LLM faults would be more realistic, collecting such data at scale is infeasible due to limited provenance information and restricted codebase access. \textit{Tricky\textsuperscript{2}} uses intentionally injected bugs as a controlled baseline, with future work aimed at incorporating naturally occurring LLM faults as such datasets become available.

\subsection{Automated Generation and Validation Pipeline}
For each TrickyBugs entry, both GPT-5 and OpenAI-oss-20b (OSS) were applied independently to the buggy and fixed versions. This produces four injected variants per problem: Human+LLM (GPT-5), Human+LLM (OSS), LLM-Only (GPT-5), and LLM-Only (OSS). Each generation is stored separately. The model output was automatically captured and stored in a language-specific directory (C++, Python, or Java). Each injection produces a modified program containing exactly one additional LLM-generated bug, while the original human-written files are preserved unchanged as the Human-only split. A language-specific validation script \footnote{Validator available at \url{https://github.com/WM-SEMERU/prj-syntax-errors}} checks for syntactic and compilation errors using \texttt{g++ -fsyntax-only}, \texttt{python -m py\_compile}, and \texttt{javac}. As illustrated in \autoref{fig:tricky2_pipeline}, this process yields five aligned variants per problem: Human-only, Human+LLM (GPT-5), Human+LLM (OSS), LLM-only (GPT-5), and LLM-only (OSS), all sharing identical test cases and fixed references. This results in a total of 11,851 buggy programs, with 3,043 per Human+LLM split and 1,361 per LLM-only split along with the original 3,043 human-written programs. A taxonomy-level distribution across the programs with LLM injected errors (n = 8,808) is shown in \autoref{fig:tricky2_taxonomy_origins}. 

\subsection{Metadata and Reproducibility}
We log model identifiers, prompt version, language, taxonomy category, and timestamps. GPT-5 used provider defaults (no fixed seed); OSS used nucleus sampling (temperature=0.7, top\_p=0.9). We release all artifacts and logs ensuring full reproducibility.

\section{Baseline Evaluation Across Splits and Tasks}

\begin{figure*}[t]
  \centering
  \includegraphics[width=0.32\textwidth]{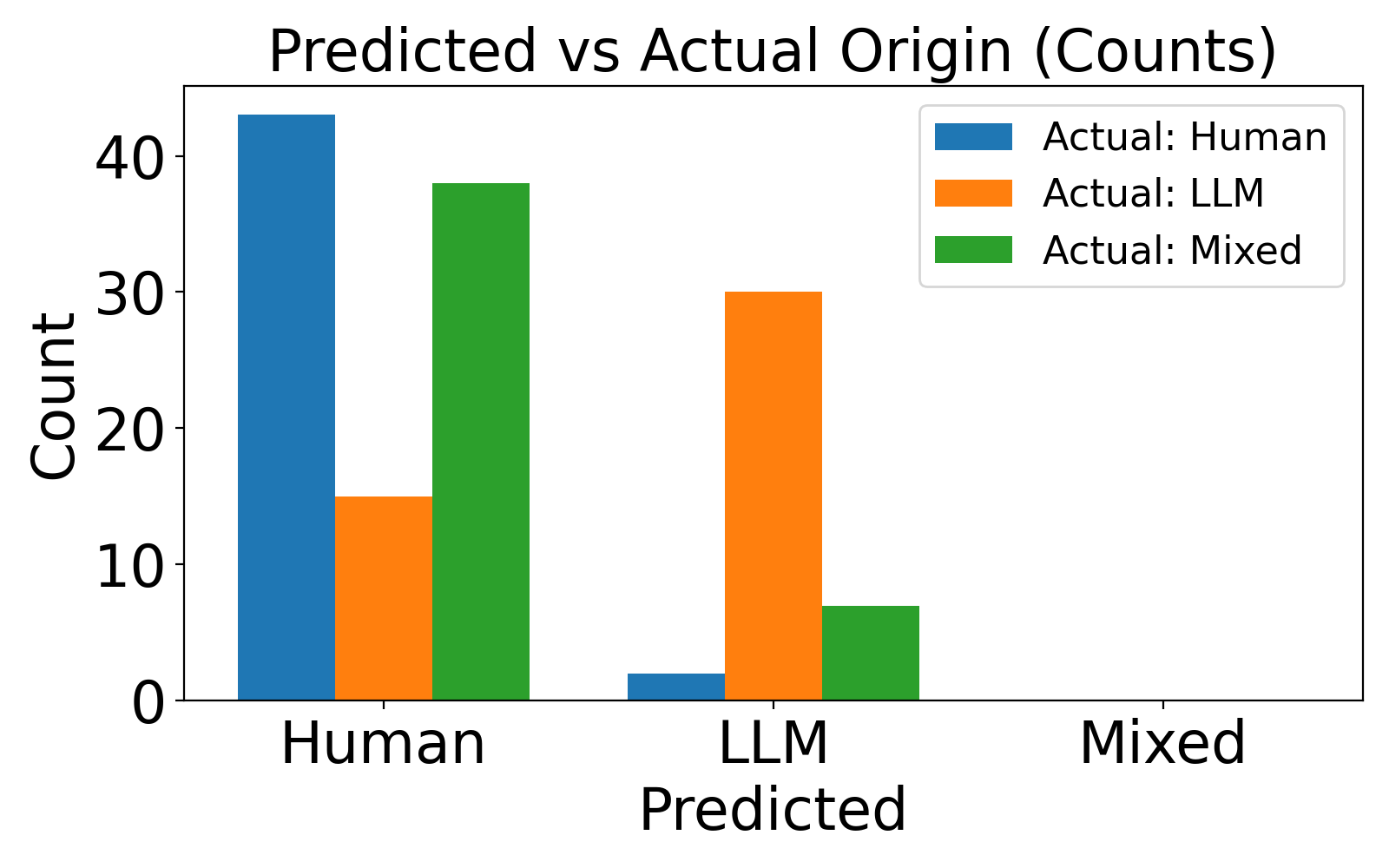}
  \includegraphics[width=0.32\textwidth]{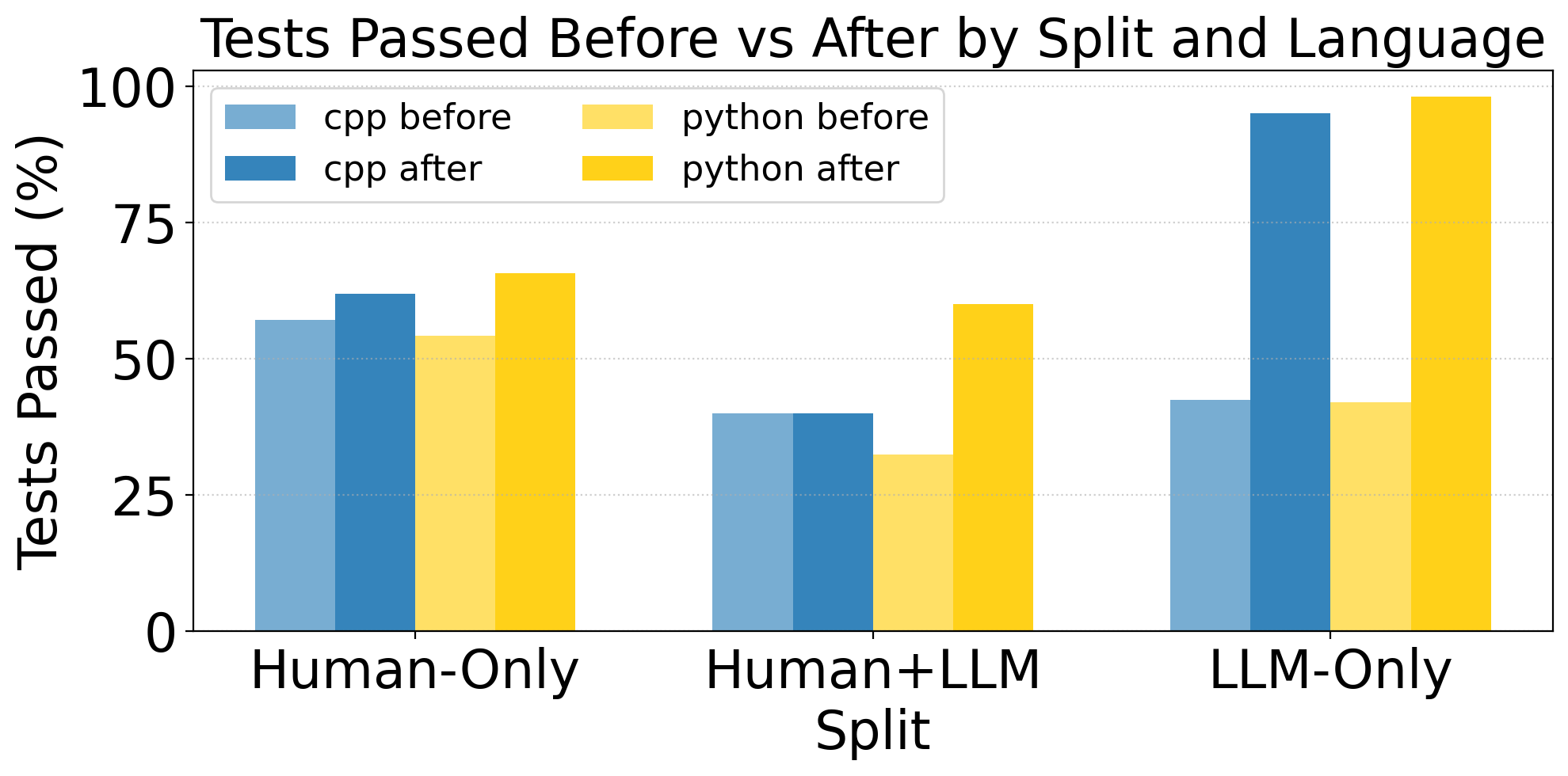}
  \includegraphics[width=0.32\textwidth]{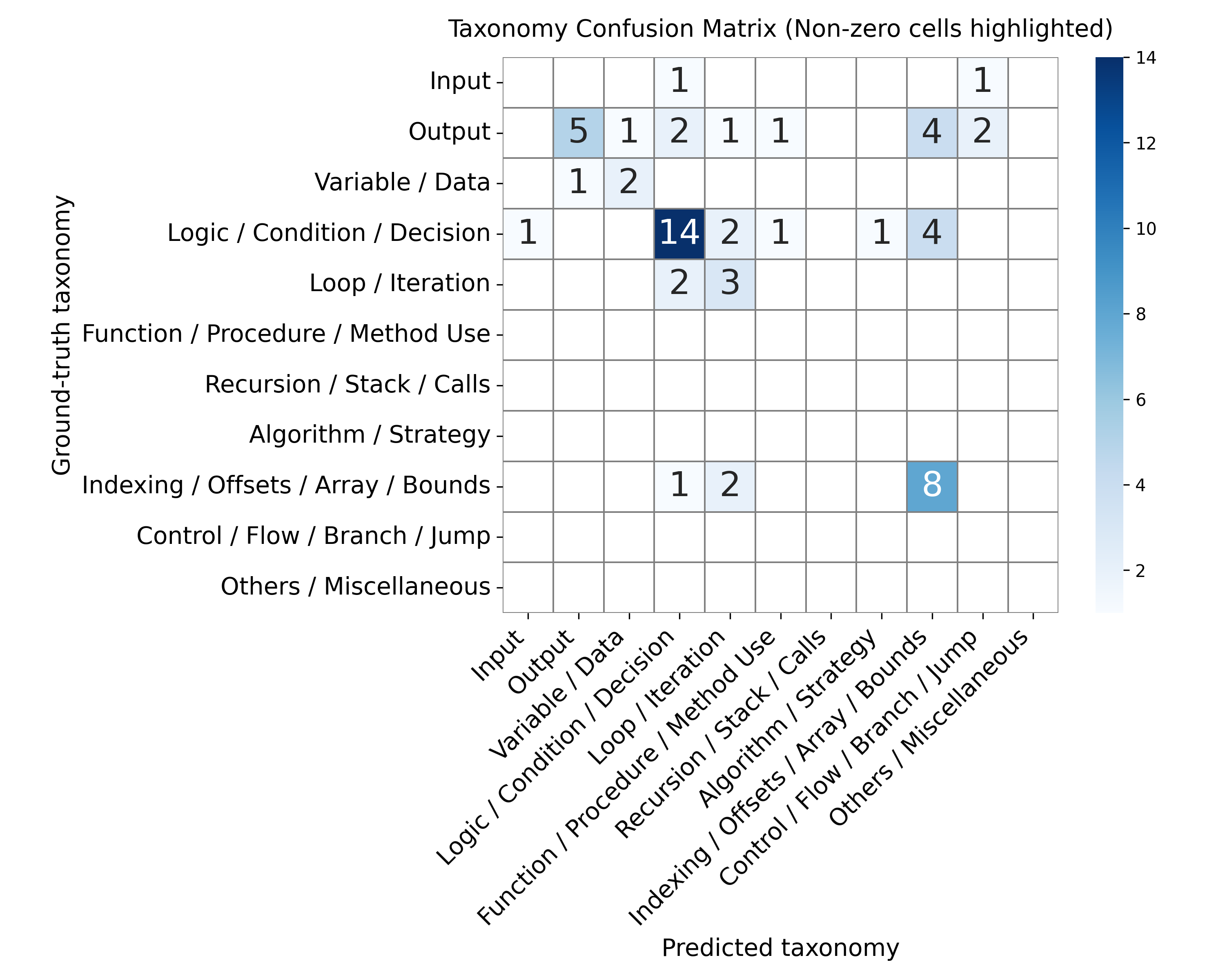}
  \Description{Three visualizations: (Left) A grouped bar chart showing predicted origin counts across human, LLM, and mixed categories for each dataset split. (Middle) A bar chart comparing repair test pass rates before and after repair for C++ and Python programs across human-only, human+LLM, and LLM-only splits. (Right) A confusion matrix heatmap showing predicted versus ground truth taxonomy levels for error identification.}
  \caption{
    \textit{Baseline evaluation visualizations:}
    (Left) predicted origin counts;
    (Middle) repair outcomes;
    (Right) taxonomy confusion heatmap.
  }
  \label{fig:baseline_bundle}
\end{figure*}

To assess basic feasibility and illustrate how \textit{Tricky\textsuperscript{2}} can be used, we conducted a small proof-of-concept baseline rather than a full empirical study. We evaluated a stratified subset of 135 difficult problems (45 per split; 15 per language across C++, Java, and Python), selected using simple difficulty heuristics (token count, number of function calls, and number of conditional branches). We prioritized more complex programs because mixed-origin defects are most likely to exhibit interaction effects in code with non-trivial control flow, where multiple bugs can mask or compound one another \cite{faultinterference, multihunkdebugging, vidziunas2024impactprogramreductionautomated}. This size provides balanced coverage across languages and error-origin splits while keeping the multi-stage evaluation pipeline, model prompting, compilation, execution, and diff analysis, computationally tractable. All tasks are evaluated with the same compiler/runtime harness and AtCoder-style tests; each run logs the model identifier, prompt template, file path, and timestamps, and we store raw model outputs (gzipped) for reproducibility.

\subsection{Task 1: Origin Classification}
The model receives only the source file and returns a JSON label in \{\texttt{human}, \texttt{LLM}, \texttt{mixed}\} with a confidence score. We report the overall accuracy and the $3 \times 3$ confusion matrix.

\subsection{Task 2: Error Identification}
The model predicts up to three character spans that contain an error with a taxonomy label. A heatmap of the predicted taxonomy level and the ground truth taxonomy level is shown in \autoref{fig:baseline_bundle}. We also release raw responses and prompt hashes to aid future analysis.

\subsection{Task 3: Program Repair}
The model returns a minimal patch that preserves the public I/O contract. We compile and run the patched program against the provided tests and report: (i) exact success rate (all tests pass), (ii) average pass-rate gain relative to baseline, (iii) total regressions (newly failing tests), and (iv) edit lines (from unified diffs). Java sources are not reported due to issues with the testing environment.

\subsection{Reporting}
All metrics are aggregated by split and language. We provide per-task summaries and per-file JSONL artifacts to support downstream replication and analysis. We release all evaluation prompts, responses, and code to encourage more thorough future empirical studies on \textit{Tricky\textsuperscript{2}}.

\section{Discussion}

Our small-scale baseline reveals several patterns that motivate further study.
First, both the \textit{human-only} and \textit{LLM-only} splits show higher repair success rates than the \textit{human+LLM} split. This suggests that mixed-origin defects introduce interaction effects that complicate reasoning and repair, even when each individual bug is comparatively simple. The absence of successful repairs in the \textit{human+LLM} C++ split further highlights this difficulty: the model failed to produce a single fixed program, despite correcting a substantial subset of human-only and LLM-only cases.

These observations reinforce the need for benchmarks that explicitly probe multi-bug and mixed-origin scenarios. Although our evaluation is limited in scope, the divergence in repair outcomes indicates that current LLMs may struggle with compounding or masking interactions between heterogeneous bug sources---a phenomenon not observable in existing datasets.

In general, these results underscore the value of \textit{Tricky\textsuperscript{2}} as a testbed for studying robustness, error interaction, and mixed-origin debugging performance.

\section{Limitations}

While \textit{Tricky\textsuperscript{2}} provides a first step toward benchmarking hybrid human+LLM software errors, several limitations remain.
First, the dataset size and domain coverage are constrained by the scope of the underlying \textit{TrickyBugs} corpus, which focuses on reasoning-oriented programming tasks rather than large-scale industrial systems.
Second, error injection relies on the behavior of specific LLMs (GPT-5 and OpenAI-oss-20b) at the time of data generation.
As these models evolve, injected error patterns may shift, introducing potential drift in future reproductions.
Third, our taxonomy, while designed to capture common programming error classes, does not yet cover semantic errors involving concurrency, memory management, or system-level APIs.
Finally, the validation stage enforces syntactic correctness but does not guarantee semantic plausibility or runnable correctness for all injected programs. These factors may introduce noise in downstream evaluations, especially for repair and error localization tasks.

\section{Conclusions \& Future Work}
\textit{Tricky\textsuperscript{2}} introduces the first controlled benchmark for studying how human and LLM-generated errors coexist, interact, and compound. By extending \textit{TrickyBugs} with systematically injected faults from GPT-5 and OpenAI-oss-20b, the dataset enables analysis of mixed-origin defects across origin classification, error identification, and automated repair tasks. Our baseline results, while limited in scope, show that hybrid human+LLM programs are more difficult to repair than human-only or LLM-only cases, highlighting interaction effects that are invisible in existing single-source benchmarks. These early findings underscore the need for evaluation settings that reflect the increasingly collaborative---and error-entangled---nature of real software development. We release \textit{Tricky\textsuperscript{2}} as a foundation for future work on robust debugging, error-aware collaboration, and the design of safer, more reliable human–AI coding workflows.

Future extensions of \textit{Tricky\textsuperscript{2}} will expand both the scale and the depth of the evaluation.
We plan to extend the corpus to additional programming languages (e.g., Rust, Go) and integrate richer static and dynamic analysis for semantic validation.
Incorporating multiple model families (e.g., Claude, Gemini, Llama) will allow comparative studies of error diversity and bias across architectures.
Beyond dataset expansion, we aim to develop agentic evaluation protocols that incorporate context gathering, execution feedback, and self-refinement, enabling multi-turn debugging experiments in hybrid human-AI settings and deeper analysis of mixed-origin error interaction phenomena.
Finally, we envision \textit{Tricky\textsuperscript{2}} as a foundation for longitudinal studies of LLM reliability tracking to determine how model-induced error profiles evolve as code-generation systems mature.

\section{Data Availability}

The \textit{Tricky\textsuperscript{2}} benchmark data are publicly available on Zenodo \cite{tricky2_dataset}, and the corresponding scripts are available on GitHub \cite{tricky2_code}.






\bibliographystyle{ACM-Reference-Format}
\bibliography{bibliography}

\end{document}